\documentclass[twocolumn]{aastex61}
\usepackage{color}

\received{December 25, 2016}
\revised{--}
\accepted{--}
\submitjournal{ApJ Letters}

\shorttitle{ALMA and Fermi observations of PKS 0637-752}
\shortauthors{Meyer et al.}

\begin{document}

\title{New ALMA and Fermi/LAT Observations of the large-scale jet of PKS~0637-752 Strengthen the Case Against the IC/CMB Model}

\correspondingauthor{Eileen T. Meyer}
\email{meyer@umbc.edu}

\author{Eileen T. Meyer}
\affil{University of Maryland, Baltimore County, 1000 Hilltop Circle, Baltimore, MD 21250, USA}

\author{Peter Breiding}
\affil{University of Maryland, Baltimore County, 1000 Hilltop Circle, Baltimore, MD 21250, USA}

\author{Markos Georganopoulos}
\affil{University of Maryland, Baltimore County, 1000 Hilltop Circle, Baltimore, MD 21250, USA}
\affil{NASA Goddard Space Flight Center, Code 663, Greenbelt, MD 20771, USA}

\author{Iv\'an Oteo}
\affil{Institute for Astronomy, University of Edinburgh, Royal Observatory, Blackford Hill, Edinburgh EH9 3HJ}
\affil{European Southern Observatory, Karl-Schwarzschild-Str. 2, 85748 Garching-bei-M\"{u}nchen, Germany}

\author{Martin A. Zwaan}
\affil{European Southern Observatory, Karl-Schwarzschild-Str. 2, 85748 Garching-bei-M\"{u}nchen, Germany}

\author{Robert Laing}
\affil{European Southern Observatory, Karl-Schwarzschild-Str. 2, 85748 Garching-bei-M\"{u}nchen, Germany}

\author{Leith Godfrey}
\affil{ASTRON, the Netherlands Institute for Radio Astronomy, Postbus 2, 7990 AA, Dwingeloo, The Netherlands}

\author{R. J. Ivison}
\affil{Institute for Astronomy, University of Edinburgh, Royal Observatory, Blackford Hill, Edinburgh EH9 3HJ}
\affil{European Southern Observatory, Karl-Schwarzschild-Str. 2, 85748 Garching-bei-M\"{u}nchen, Germany}



\begin{abstract}
The \emph{Chandra} X-ray observatory has discovered several dozen
anomalously X-ray-bright jets associated with powerful quasars.  A
popular explanation for the X-ray flux from the knots in these
jets is that relativistic synchrotron-emitting electrons
inverse-Compton scatter Cosmic Microwave Background (CMB) photons to
X-ray energies (the IC/CMB model).  This model predicts a high
gamma-ray flux which should be detectable by the \emph{Fermi} Large
Area Telescope (LAT) for many sources. GeV-band upper limits from
\emph{Fermi}/LAT for the well-known anomalous X-ray jet in
PKS~0637$-$752 were previously shown in \cite{meyer2015} to violate
the predictions of the IC/CMB model.  Previously, measurements of the
jet synchrotron spectrum, important for accurately predicting the
gamma-ray flux level, were lacking between radio and infrared
wavelengths.  Here we present new Atacama Large
Millimeter/submillimeter Array (ALMA) observations of the large-scale
jet at 100, 233 GHz, and 319 GHz which further constrain the
synchrotron spectrum, supporting the previously published empirical
model.  We also present updated limits from the \emph{Fermi} Large
Area Telescope (LAT) using the new `Pass 8' calibration and
approximately 30\% more time on source. With these deeper limits we
rule out the IC/CMB model at the 8.7$\sigma$ level.  Finally, we
demonstrate that complete knowledge of the synchrotron SED is critical
in evaluating the IC/CMB model.

\end{abstract}




\section{Introduction}
The radio-loud quasar PKS~0637$-$752 ($z$=0.651) was the first astrophysical
source observed by the \emph{Chandra} X-ray observatory, shortly after
its launch in 1999. This source also gave us \emph{Chandra's} first
scientific discovery, as it unexpectedly detected a kpc-scale X-ray
jet associated with PKS~0637$-$752, with a very high X-ray flux level
and hard ($\alpha_x=0.85\pm0.1$) spectrum which precluded it being the
tail of the radio-optical synchrotron spectrum
\citep{chartas2000,schwartz2000}. Nearly twenty years later,
\emph{Chandra} has now discovered several dozen more `anomalous' X-ray
jets, typically in powerful quasar sources, where the hard-spectrum
X-ray flux is clearly from a second component, distinct from the
radio-optical synchrotron spectrum which turns over before the
infrared \citep[see, e.g.,][for reviews]{harris2006,hardcastle2006}.

In the initial discovery papers on PKS~0637$-$752,
\cite{schwartz2000} and \cite{chartas2000} ruled out thermal brehmstrahlung, synchrotron
self-Compton, and inverse Compton off the Cosmic Microwave Background (IC/CMB) from a mildly
relativistic flow as possible explanations for the anomalous X-ray
flux in the jet, leaving open the possibility of a second synchrotron
spectrum from an apparently co-spatial second electron energy
distribution. \cite{tavecchio2000} and \cite{celotti2001}
independently revisited the IC/CMB mechanism, finding that it could be
consistent with the X-ray flux if the kpc-scale jet remains highly
relativistic, with bulk Lorentz factors $\Gamma\sim$10, and the jet is
pointed within a few degrees of our line-of-sight. This version of the
IC/CMB model has become by far the most popular explanation for the
anomalous X-ray jets, always requiring small viewing angles and
high bulk $\Gamma$ values
\citep[e.g.,][]{sambruna2004,jorstad2006,miller2006,marshall2011,perlman2011,godfrey2012,kharb2012,stanley2015}.

\begin{figure*}[ht]
\vspace{20pt}
\begin{center}
\includegraphics[width=6.5in]{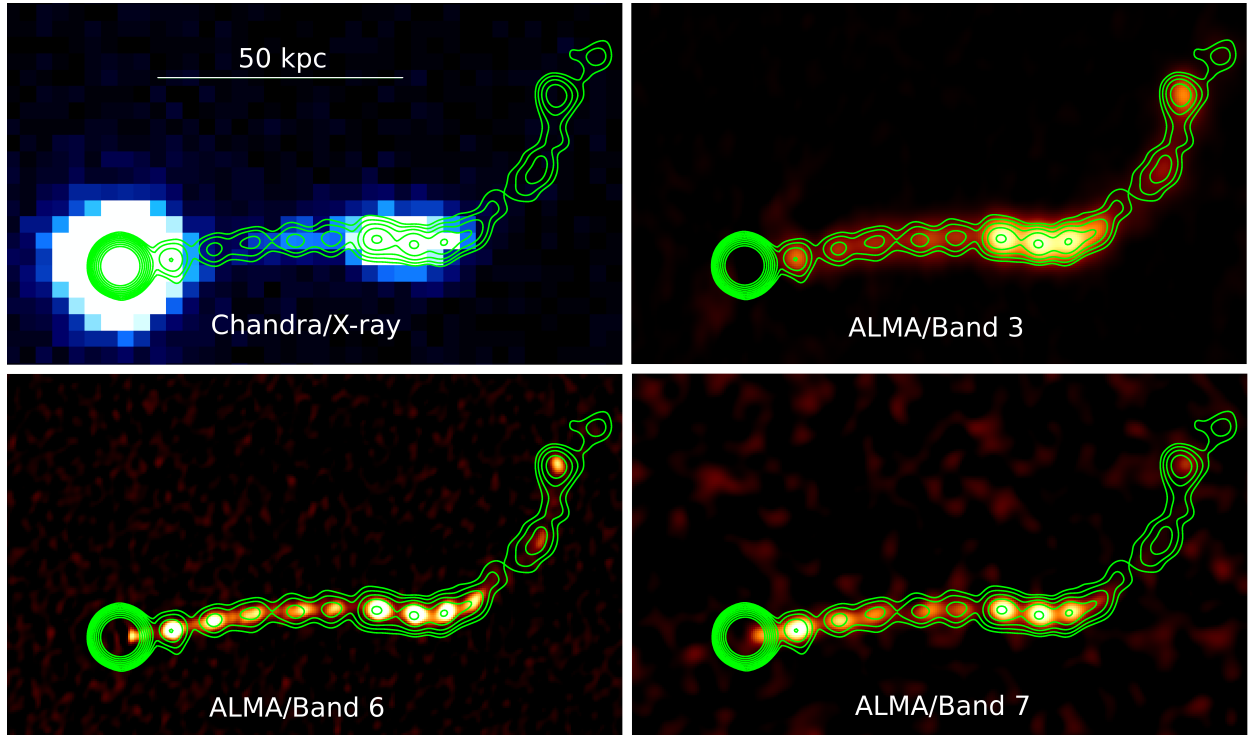}
\end{center}
\caption{\label{fig:imaging} At top left, an archival \emph{Chandra}
  X-ray image of PKS~0637$-$752 with 17~GHz radio contours
  \citep{godfrey2012} overlaid. The same contours are shown in all
  four images. The other panels show the ALMACAL program images
  (uncorrected for primary beam) for PKS~0637$-$752 in bands 3, 6, and
  7. In the ALMA imaging the core has been subtracted to allow for
  higher contrast in the fainter knots. Primary beam correction was
  applied before measuring individual knot fluxes. 
}
\end{figure*}

The IC/CMB model has come under criticism for several reasons,
including unrealistically high jet power requirements, inability to
explain jet-to-counterjet ratios in less highly beamed members of the
parent source population, difficulty matching the `knotty' jet
structure, and unusually long jet lengths implied by the small viewing
angle
\citep{dermer2004,jorstad2004,uchiyama2006,hardcastle2006,hardcastle2016}. However,
until recently there was no clear test of the IC/CMB model. The central difficulty is that both
IC/CMB and a double-synchrotron models are able to reproduce the
observed radio to X-ray SEDs of large-scale-jets equally well
\citep[e.g., see direct comparison in ][]{cara2013}.

\citet[][]{georganopoulos2006}, hereafter G06 suggested that the
\emph{Fermi}/LAT would be able to break the degeneracy between these
two models, by looking for the required high level of gamma-ray flux
under the IC/CMB interpretation, which is obviously not expected under
a second-synchrotron scenario. The \emph{Fermi} test is extremely
powerful because there are no free parameters at all involved in the
predicted gamma-ray flux level: the IC/CMB spectrum is essentially an
exact copy of the synchrotron spectrum, shifted in frequency and
luminosity proportionally to $B/\delta$ and $(B/\delta)^2$
respectively (where $B$ is the magnetic field strength and $\delta$
the Doppler factor), and hence the requirement to set the
normalization of the IC/CMB spectrum to match the X-ray flux
completely fixes the gamma-ray flux level. An upper limit \emph{or} a
detection by \emph{Fermi} significantly below the expected level at
GeV energies would clearly rule out an IC/CMB origin for the anomalous
X-rays. 

In recent work, we have used upper limits from \emph{Fermi} to rule
out IC/CMB as the source of the kpc-scale X-rays at the $>$99.99\%
level in 3C~273, and at the 99.98\% level in PKS~0637$-$752
\citep[][hereafter M15]{meyer2014,meyer2015}. Although \emph{Fermi}
lacks the spatial resolution to detect the large-scale jet separately
from the gamma-ray bright cores in these sources, we were able to
derive deep upper limits using times when the cores were quiescent.
In this letter, we present new ALMA observations and updated limits
from \emph{Fermi}/LAT which show that our original claims can be
strengthened, and that IC/CMB is now ruled out in PKS~0637$-$752 at
the 8.7$\sigma$ level.

\section{Data Analysis}
We describe here new observations obtained from ALMA and the \emph{Fermi}/LAT for PKS~0637$-$752. All other data are taken from M15 and references therein.  
\subsection{ALMA}

PKS\,0637$-$752 is a bright southern source which is routinely used as
an ALMA calibrator. The ALMA data used in this paper have been taken
from ALMACAL, a wide and deep submm survey which is being carried out
using ALMA calibration observations \citep{Oteo2016a,Oteo2016b}. We
refer the reader to \cite{Oteo2016a} for details on data extraction
and calibration. Briefly, for each scheduling block of a given science
project, we run the ALMA pipeline and extract the calibrated
visibilities of all calibrators used. Then a run of self-calibration
is carried out (always possible thanks to the brightness of the
calibrators) and the dominant point-source component at the phase
centre is subtracted from the calibrated visibilities. This
subtraction is necessary to combine data taken in different
configurations (and therefore different times) due to the variability
of the core.  Finally, for a given calibrator, we combine all
available visibilities on each band to create a deep image (natural
weighing is used during imaging). In the current phase of ALMACAL,
there is available data for PKS\,0637$-$752 in ALMA bands 3 ($\sim
3\,{\rm mm}$), 6 ($\sim 1.2 \, {\rm mm}$) and 7 ($\sim 870 \, {\rm \mu
  m}$). The sensitivity and beam size of each map (shown in
Figure~\ref{fig:imaging}) are: $\sigma_{\rm B3} = 66.2 \, {\rm \mu Jy
  \, beam^{-1}}$ and $\theta_{\rm B3} = 1.03 \times 0.79''$,
$\sigma_{\rm B6} = 21.6 \, {\rm \mu Jy \, beam^{-1}}$ and $\theta_{\rm
  B6} = 0.38 \times 0.29''$, and $\sigma_{\rm B7} = 46.0 \, {\rm \mu
  Jy \, beam^{-1}}$ and $\theta_{\rm B7} = 0.33 \times 0.22''$.

\begin{figure}[t]
\vspace{20pt}
\begin{center}
\includegraphics[width=3.5in]{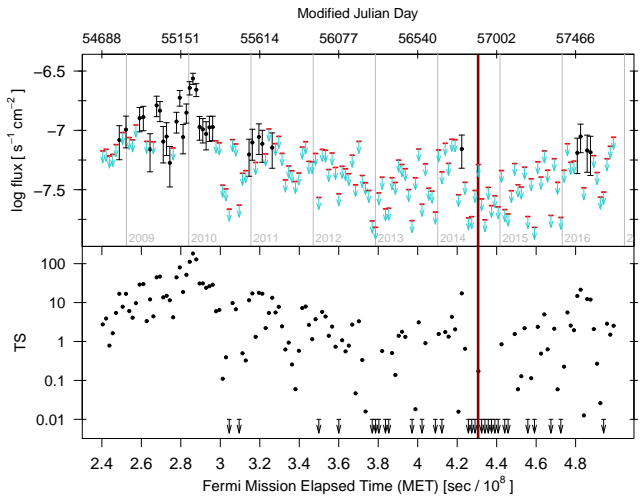}
\end{center}
\caption{\label{fig:lightcurve} \emph{Upper Panel:} The full
  lightcurve of PKS~0637$-$752 from the onset of \emph{Fermi}/LAT
  observations in August 2008 to November 2016 (beginning of calendar
  years shown by light gray vertical lines). Data were binned in time
  to produce bins of equal good time interval (GTI) time on source
  totaling 1 week, corresponding to roughly 2.8 weeks in real time.
  The fluxes shown are for the full energy band from
  100~MeV$-$100~GeV, where fluxes with error bars are shown when the
  TS was greater than 10, and upper limits are shown when the source
  TS was less than 10. The total \emph{Fermi}/LAT flux shown here is
  clearly dominated by the variable core. \emph{Bottom Panel:} The
  corresponding TS for the same bins shown in the upper panel. In both
  panels the dark red line corresponds to the time cut-off of the M15
  study.}
\end{figure}

We measured the fluxes for the X-ray brightest knot wk8.9 and the
cluster of four X-ray bright knots in the ALMA band 3, 6 and 7
primary-beam-corrected maps using contours around each knot using
\texttt{casaviewer}, as the knots are resolved (with total fluxes
about twice that of the peak). For wk8.9, we measured fluxes in the
three bands of 1.9, 3.2, and 2.6 mJy, respectively.  For the combined
knots the values are 6.0, 10.7, and 7.9 mJy. For all measurements, the
error on the flux is dominated by the error on the absolute flux scale
calibration; current estimates suggest 10\% as the maximum value,
which we use for all ALMA fluxes presented in this paper.

\subsection{Fermi}
To get the deepest possible upper limits (or lowest detected flux
level) for the gamma-ray output of the combined PKS~0637$-$752 core
and jet, we utilize the `progressive-binning' approach first described
in \cite{meyer2014}, which can be consulted for further details as
well as M15. The method involves first making a lightcurve for
PKS~0637$-$752, which is a moderately bright \emph{Fermi}/LAT source
with a $\sqrt{TS}$ of 22 (where TS is the Test Statistic value,
roughly significance squared) in the 3FGL \emph{Fermi} four-year
point-source catalog \citep{4fgl}. The bins were selected to have good
time intervals (GTIs) totaling 1 week on source, corresponding to
approximately 2.8 weeks in real time, for a total of 155 bins. Before
making the light curve, we analyzed a 10 degree region around
PKS~0637$-$752 for potential new sources not accounted for in the
latest 3FGL catalog, using data taken between 239557417 and 500455771
MET, corresponding to calendar dates 2008 August~4 to
2016~November~10. We iteratively fit for the sky position and spectral
parameters for any new source (as shown by a TS residual over
$\sim$10) using a maximum-likelihood approach as described in M15. We
added 13 new sources to the base model populated with 3FGL catalog
sources supplied by the \texttt{make3FGLxml} tool, with the nearest
being a TS=19 source at 3.45 degrees from PKS~0637$-$752.

With our updated source model, we then made the light curve shown in
Figure~\ref{fig:lightcurve}. Since the cut-off time corresponding to
the M15 study (vertical red line), almost 26 months have passed in
which the PKS~0637$-$752 core has remained relatively quiescent. We
ordered the 155 light curve bins according to TS and then
progressively combined them in our analysis, combining first the
lowest two, then lowest three, and so on. At each step in the combined
binning, where one additional bin is included in to the analysis, we
compute the maximum-likelihood flux for PKS~0637$-$752 using a
power-law model with a fixed spectral index of 2.7\footnote{Results
  are not significantly altered by using a free spectral index, and we
  found no sign of a significant spectral change from an index of
  approximately 2.7 over the entire course of the progressive binning
  analysis when the spectral index was free.}. At each step we also
compute the flux or 95\% upper limit in five individual energy bands
corresponding to 0.1-0.3~GeV, 0.3-1~GeV, 1-3~GeV, 3-10~GeV, and
10-100~GeV, matching those used in the published \emph{Fermi}
catalogs.

\begin{figure}[t]
\vspace{20pt}
\begin{center}
\includegraphics[width=3.5in]{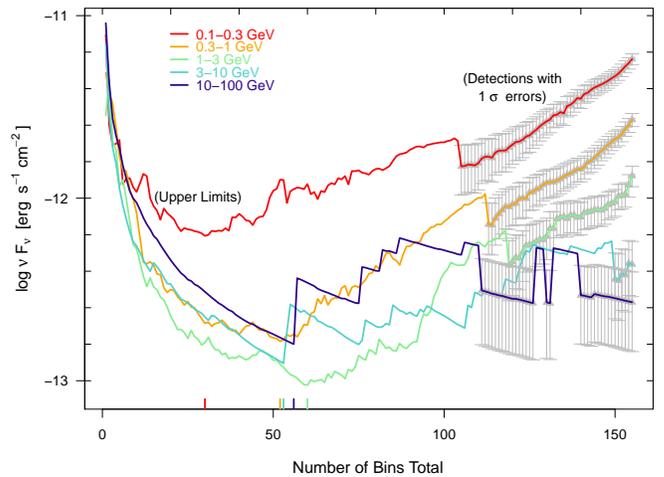}
\end{center}
\caption{\label{fig:newcurves} Here we show the results of the progressive binning analysis, as upper limits versus total bins added.  After ordering the lightcurve bins from lowest to highest TS (using photons from 100 MeV to 100 GeV), we progressively binned 1, 2, ... 155 total bins, in order to increase the time on source while avoiding times when the core was in a high flux state.  As shown, the typical behavior in all energy bins is to show decreasing upper limits with time, reaching a minimum, and then increasing up to the point that the source reached a test significance greater than 10 in that band (points shown with error bars).  We use the minimum upper limit to determine the absolute upper limit on the large scale jet in each band. }
\end{figure}

\begin{figure*}[t]
\begin{center}
\includegraphics[width=5.5in]{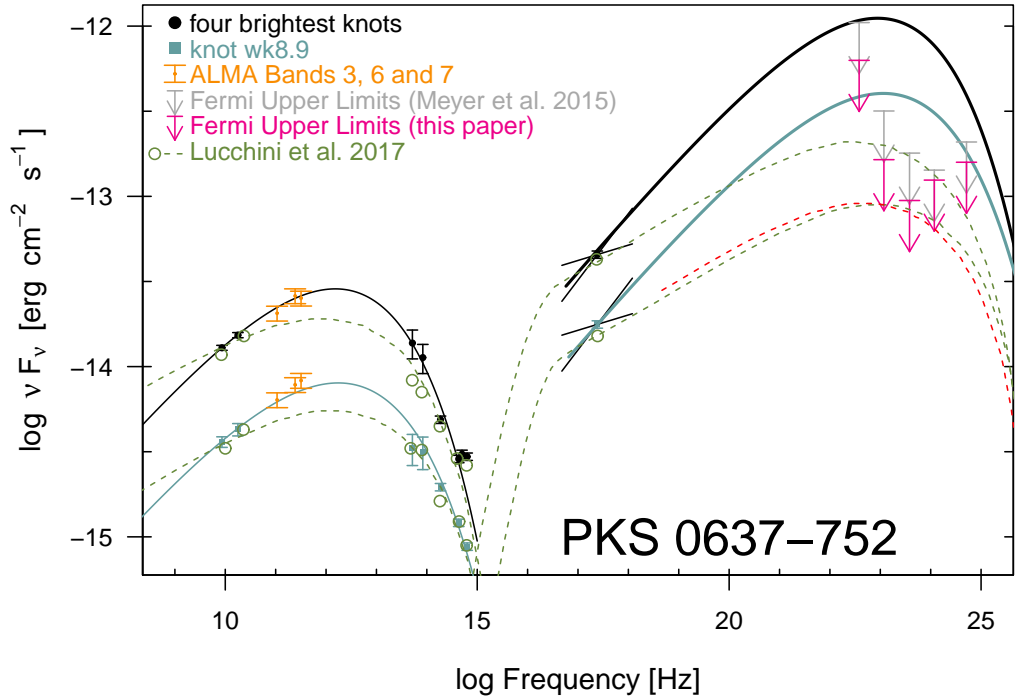}
\end{center}
\caption{\label{fig:mainresult} Here we show the SEDs for the four
  brightest knots (black points and lines) and the X-ray brightest
  know wk8.9 (cyan points and lines).  The new ALMA fluxes are plotted
  in dark yellow. The green open circles and dashed curves are taken
  from L17 -- note the discrepancy with the observed
  radio, ALMA, and IR fluxes.  At upper right, the new Fermi upper limits
  are plotted in magenta (previous limits from M15 are shown in grey
  for comparison). The new upper limits rule out the IC/CMB model curves
  (thick black and cyan lines) at an 8.7$\sigma$ level for the
  combined four brightest knots and at the 5.3$\sigma$ level for the
  single knot wk8.9. The red dashed curve is the result of using the simple shift formula on the synchrotron curve of L17 for knot wk8.9. }
\end{figure*}

In Figure~\ref{fig:newcurves}, we show the results of the progressive
binning analysis as 95\% upper limits versus time, up until a flux with
TS=10 is detected, at which point we show detections with error
bars. The lowest limits in each of the five energy bands were reached
after 31, 52, 60, 53, and 56 total bins were combined, respectively.
The corresponding 95\% upper limit fluxes, plotted in red in
Figure~\ref{fig:mainresult}, are listed in Table~\ref{table:limits}.

\clearpage
\section{Results and Discussion}

\subsection{Ruling out IC/CMB with the New Data}

We present in Figure~\ref{fig:mainresult} the updated SED for the four
brightest knots (black lines) and for the single X-ray brightest knot
wk8.9 (cyan lines). The new ALMA data have been overplotted in dark
yellow; note that they fall nearly exactly on the synchrotron model
curves which are unchanged from M15. The radio and IR/optical data
shown as solid points are also unchanged from the previous paper.

With the addition of the ALMA data points, we can now say that the
synchrotron SEDs are quite well-constrained, as the data gap between
ALMA and IR frequencies is comparable to the spectral width of
synchrotron emission from a monoenergetic electron population and the
actual flux there cannot deviate significantly from our empirical SED.
 The empirical (by eye) fit is simply a power-law with a scaled exponential cutoff,

\begin{equation}
  \nu f_\nu=N\left(\frac{\nu}{10^{10}\mathrm{\,Hz}}\right)^\gamma \exp{\left(-\left(\frac{\nu}{\nu_1}\right)^\beta\right)}
\end{equation}

 For the four knots, the parameter values are $N=2\times
 10^{-14}$~erg~s$^{-1}$~cm$^{-2}$, $\gamma=0.35$, $\nu_1=4\times
 10^{11}$~Hz, and $\beta=0.25$.  For the single knot the values are
 $N=6.3\times 10^{-15}$~erg~s$^{-1}$~cm$^{-2}$, $\gamma=0.35$,
 $\nu_1=2.\times 10^{11}$~Hz, and $\beta=0.23$. The radio index of the
 empirical model between 8.6 and 17.8 GHz is 0.245, consistent with
 that based on the two ATCA radio points alone, which yields
 $\gamma$=0.24$\pm$0.05. Based on the empirical fits to the
 synchrotron spectra shown, we have produced corresponding (empirical)
 IC/CMB model curves, shown as thicker black and cyan lines from X-ray
 to TeV energies using the shift formulae given in G06.

Recalling that the requirement that the IC/CMB curves match the
observed X-ray flux completely fixes the GeV-band prediction, we can
now examine the upper limits from \emph{Fermi}. The previous M15
limits are shown in gray, while the new, deeper limits from the most
recent analysis are shown in magenta. Clearly, the IC/CMB model is
ruled out at a very high level of significance. Taking the predicted
model curves shown in Figure~\ref{fig:mainresult} for the IC/CMB
spectrum, we can calculate the significance of our non-detection in
each energy band using the profile likelihood method. With $L$ equal
to the logarithm of the likelihood,the statistic $2\Delta L$ is
distributed as a $\chi^2$ with one degree of freedom when all other
parameters are fixed. The value of the normalization of the source is
increased until the flux equals the predicted value, and we record the
resulting $2\Delta L$ value, which is converted into a percent
probability and sigma value for a one-sided limit using standard
$\chi^2$ tables. For the four brightest knots in the jet combined, the
deepest single-band limit is in band 3, where our upper limit implies
that IC/CMB is ruled out at the 5.6$\sigma$ level.  Overall, IC/CMB is
ruled out at a 8.7$\sigma$ limit when we combine the individual band
results using the inverse normal method. For the single brightest
knot, the deepest single limit is also in band 3, equivalent to a
99.97\% or 3.5$\sigma$ upper limit. The combined results of all bands
using the inverse normal method implies an overall 5.3$\sigma$ upper
limit for the single X-ray brightest knot.

Because of jet one-sidedness, we know that the large-scale jet must be
at least mildly relativistic, and thus some level of IC/CMB emission
will be produced, even if it is not responsible for the anomalous
X-rays.  We can use the updated \emph{Fermi} limits to put an upper
limit on the value of $\delta$ assuming the magnetic field is in
equipartition, with $B\delta=1.35\times10^{-4}$ G (where we have
assumed a knot radius of 3.5 kpc, a spectral index of $\alpha=0.7$ and
$\gamma_\mathrm{min}$ = 100). For the 95\% limits shown in
Figure~\ref{fig:mainresult}, the limit for the four brightest knots is
$\delta<5.3$.

\begin{deluxetable*}{ccccccclcl}[t]
\tabletypesize{\scriptsize}
\tablecaption{Results of the Fermi Data Analysis}
\centering
\tablehead{
Band & $E_1$ & $E_2$ & log Freq. & 95\% Limit              & Bins & \multicolumn{2}{c}{Combined Knots\tablenotemark{*}} & \multicolumn{2}{c}{Knot wk8.9} \\
& (GeV) & (GeV) &   (Hz)    & (erg s$^{-1}$ cm$^{-2}$)  & Added &  predicted $F_\mathrm{IC/CMB}$ & \% Ruled & predicted $F_\mathrm{IC/CMB}$ & \% Ruled\\
      &       &       &           &                         & & (erg s$^{-1}$ cm$^{-2}$) & Out  & (erg s$^{-1}$ cm$^{-2}$) &  Out    \\
(1)    & (2)  & (3)   & (4)   & (5)       & (6) & (7) & (8) & (9) & (10) \\
}
\startdata
1 & \phantom{00}0.1           & \phantom{00}{0.3}           & 22.6 & 6.29$\times 10^{-13}$ & 31 & 1.1$\times 10^{-12}$ & 98.99  & 3.8$\times 10^{-13}$ & 88.9  \\
2 & \phantom{00}0.3           & \phantom{00}{1}\phantom{00} & 23.1 & 1.64$\times 10^{-13}$ & 52 & 1.1$\times 10^{-12}$ & 99.99999  & 4.0$\times 10^{-13}$ & 99.7 \\
3 & \phantom{00}1\phantom{00} & \phantom{00}{3}\phantom{00} & 23.6 & 9.45$\times 10^{-14}$ & 60 & 1.0$\times 10^{-12}$ & $>$99.99999 & 3.8$\times 10^{-13}$ & 99.97 \\
4 & \phantom{00}3\phantom{00} & \phantom{0}{10}\phantom{00} & 24.1 & 1.25$\times 10^{-13}$ & 53 & 7.7$\times 10^{-13}$ & 99.998 & 3.1$\times 10^{-13}$ & 99.5 \\
5 & \phantom{0}10\phantom{00} &            100\phantom{00}  & 24.7 & 1.58$\times 10^{-13}$ & 56 & 3.9$\times 10^{-13}$ & 98.8  & 1.8$\times 10^{-13}$ & 96.1  \\
\enddata
\label{table:limits} 

\tablenotetext{*}{Combined Knots are wk7.8, wk8.9, wk9.7, and wk10.6.}

\end{deluxetable*}

\subsection{The Importance of a Complete Synchrotron SED \label{importanceofSED}}
In the IC/CMB scenario, because the X-ray to GeV SED is a copy of the
synchrotron, it is critical to ensure accuracy and completeness in the
synchrotron spectrum to the extent possible. As an example of this, in
Figure \ref{fig:mainresult} we show as green open circles and dashed
lines the data and model fits recently presented in \citet[][hereafter
  {L17}]{lucchini2016} where it is argued that the IC/CMB model is not
ruled out for PKS~0637$-$752, based on the M15 limits shown in
gray.\footnote{We note that the L17 data set does not take into
  account the updated infrared fluxes presented previously in M15,
  resulting in a discrepancy at those wavelengths for the combined
  knots.}. However, the radio -- sub-mm model spectra of L17 are
significantly steeper (or more flat in $\nu F_\nu$) than the actual
data require, substantially underproducing the ALMA fluxes. This is
critical, as this part of the synchrotron spectrum determines the SED
slope below the peak and therefore, the rise of the high energy energy
component from the X-ray to GeV energies. In other words, adopting a
steeper-than-observed radio spectrum reduces the anticipated GeV
emission for a given level of X-rays and can lead to incorrect
conclusions for the viability of the IC/CMB model. The radio spectral
index at the low radio frequencies ($<<$1 GHz) corresponding to the
same electrons that produce the X-rays is $\alpha_r$=0.65 in our
empirical model, compared to a value of approximately 0.87 used in
L17. The latter value is close to the X-ray spectral index value of
$\alpha_x$=0.85$\pm$0.1 reported in \cite{chartas2000}, as required
under the standard IC/CMB model. It should be noted that, in the
context of the IC/CMB model, the inferred radio index is already in
some tension with the X-ray spectral index, as seen in
Figure~\ref{fig:mainresult}.


\subsection{Physical Versus Empirical Models for Testing IC/CMB \label{numerical}}

We now address the question of whether models that include some
physical considerations such as electron energy losses can alter our
conclusions. This was suggested by L17, in particular that radiative
losses steepen the high-energy end of the electron energy
distribution, resulting to suppressed GeV emission. The problem with
this argument is that there is a single electron energy distribution
that is required to produce both the synchrotron and the IC/CMB
components, and the steepness of the high-energy tail of the EED is
uniquely determined by the IR-optical \emph{observed spectrum}, so
that the synchrotron SED determines the IC/CMB SED. Thus, regardless
of the way one fits the synchrotron spectrum, the IC/CMB SED will be a
copy of the synchrotron one with the degree of shifting set by the
requirement that the IC/CMB reproduces the X-rays (G06). To
demonstrate this, we have shifted the L17 numerical synchrotron SED of
knot wk8.9 according to the expressions of G06 so that it reproduces
the X-ray emission of the knot. The result is plotted in Figure
\ref{fig:mainresult} as a dashed red line. As expected, this line is
extremely close to the numerical IC/CMB result of L17.






\section{Summary \& Conclusions}
We have presented new ALMA observations of the anomalously
X-ray-bright quasar jet PKS~0637$-$752 at 100 and 233 GHz. These
observations confirm that the bright knots of this source have a
synchrotron peak at approximately $10^{13.5}$ Hz and are consistent
with the empirical model presented in M15.
We have presented new \emph{Fermi} upper limits on PKS~0637$-$752 for
the large-scale jet using the progressive binning method.  These
limits improve on those previously reported in M15 through both more
time on source, and the use of `pass 8' calibration for the
\emph{Fermi} data.  The updated limits represent a combined
8.7$\sigma$-level ruling out of the standard IC/CMB model, which we
find does not produce the required gamma-ray emission. A recent claim
for the opposite (L17) stems from using a dataset that did not include critical ALMA observations and from using an incorrect radio spectral index (see \S
\ref{importanceofSED}).  We also note that for the purpose of
evaluating the standard IC/CMB model, an empirical description of the
synchrotron SED is equivalent to physical models (see \S
\ref{numerical}).

\acknowledgments

E.T.M and M.G. acknowledge NASA Grant ADAP14-0122 and NASA Fermi
grant 81051.

I.O., and R.J.I. acknowledge support from the European Research
Council in the form of the Advanced Investigator Programme, 321302,
{\sc cosmicism}.

L.G. acknowledges funding from the European Research Council under the
European Union’s Seventh Framework Programme (FP/2007-2013) / ERC
Advanced Grant RADIOLIFE-320745.

This paper makes use of the following ALMA data: 2012.1.00075.S,
2012.1.00335.S, 2012.1.00554.S, 2012.1.00603.S, 2012.1.00641.S,
2012.1.01108.S, 2013.1.00063.S, 2013.1.00070.S, 2013.1.00280.S,
2013.1.00287.S, 2013.1.00319.S, 2013.1.00450.S, 2013.1.00535.S,
2013.1.00700.S, 2013.1.00993.S, 2013.1.01136.S, 2013.1.01155.S,
2015.1.00190.S, 2015.1.00204.S, 2015.1.00979.S. ALMA is a partnership
of ESO (representing its member states), NSF (USA) and NINS (Japan),
together with NRC (Canada), NSC and ASIAA (Taiwan), and KASI (Republic
of Korea), in cooperation with the Republic of Chile. The Joint ALMA
Observatory is operated by ESO, AUI/NRAO and NAOJ. The National Radio
Astronomy Observatory is a facility of the National Science Foundation
operated under cooperative agreement by Associated Universities, Inc.

%

\vspace{5mm}
\facilities{VLA, ALMA, HST(STIS,ACS,WFC3), Chandra, Fermi}


\software{CASA \citep{2012ASPC..461..849P}
          }

\end{document}